\documentclass[letterpaper, paper,11pt]{AAS}		% for final proceedings (20-page limit)

\usepackage{bm}
\usepackage{amsmath}
\usepackage[colorlinks=true, pdfstartview=FitV, linkcolor=black, citecolor= black, urlcolor= black]{hyperref}
\usepackage{overcite}
\usepackage{footnpag}			      	% make footnote symbols restart on each page

\usepackage{subcaption}
\usepackage{amssymb,amsfonts}

\PaperNumber{26-754}

\begin{document}

% Fix corrupted/incompatible \@tabular definition causing
% "Missing $ inserted" on every tabular environment.
\makeatletter
\def\@tabular{%
  \leavevmode
  \hbox\bgroup$%
  \let\@acol\@tabacol
  \let\@classz\@tabclassz
  \let\@classiv\@tabclassiv
  \let\\\@tabularcr
  \@tabarray
}
\makeatother

\title{DESIGNING DENSE SATELLITE CLUSTERS FOR DISTRIBUTED SPACE-BASED DATACENTERS}

\author{Jules Pénot\thanks{Graduate Student, Department of Aeronautics and Astronautics, Massachusetts Institute of Technology, 77 Massachusetts Ave, Cambridge, MA 02139.}
\ and Hamsa Balakrishnan\thanks{William E. Leonhard (1940) Professor, Department of Aeronautics and Astronautics, Massachusetts Institute of Technology, 77 Massachusetts Ave, Cambridge, MA 02139.}
}

\maketitle{}

\begin{abstract}

Recent proposals for datacenters in sun-synchronous Low Earth Orbit (LEO) rely on a large number of compute satellites formation-flying in dense clusters. Designing such satellite clusters requires optimizing the satellites’ orbital geometry under several safety and operational constraints applied throughout the cluster’s entire orbit. These constraints include guaranteeing a minimum inter-satellite spacing, unobstructed solar power for every satellite, and that each satellite have a stable set of nearest neighbors with which it can maintain inter-satellite links (ISLs).

In this work, we propose two main cluster orbital designs, parametrized by the minimum inter-satellite spacing $R_{\min}$ and the cluster radius $R_{\max}$: a planar cluster, and a 3D cluster. We show by construction and numerical analysis that both cluster orbital designs are consistent with the inter-satellite spacing, unobstructed sun-vector, and inter-satellite line of sight constraints. The proposed planar architecture is the most efficient packing of satellites in a plane for given $R_{\min}$ and $R_{\max}$ values, and our 3D architecture allows for the number of datacenter satellites to scale proportionally to $(R_{\max}/R_{\min})^3$, an improvement over all previous LEO datacenter cluster designs.

Finally, for a given satellite cluster, we formulate and solve an integer optimization problem that maps a VL2-like Clos network datacenter switching fabric onto the satellites and their corresponding set of feasible inter-satellite links. We confirm that both the planar and 3D architectures allow for sufficiently many permanently unobstructed ISLs to replicate the switching fabric of terrestrial datacenters within the cluster. We also examine the tradeoff between the number of ISLs each satellite can simultaneously sustain, and the corresponding number of cluster satellites that must be dedicated as aggregation and intermediate switches. 

\end{abstract}

\section{Introduction}

Recent trends suggest that power availability will soon become the main factor limiting the construction and scaling of new datacenters on Earth. As a result, the plans for new datacenter projects increasingly include the construction of power sources such as solar farms, gas turbines, and even nuclear power plants \cite{oltean_why_2024}. Space-based solar power has several advantages over terrestrial solar farms: most notably, solar panels in dawn-dusk sun-synchronous orbit (SSO) receive solar flux throughout their entire orbit, year-round, making them approximately five times as productive as equivalent panels on Earth. However, transmission losses are too large for the effective use of space-based solar power to power terrestrial datacenters \cite{mizrahi_space_2025}. This has motivated a series of recent proposals to place datacenters directly in low Earth orbit (LEO), in proximity to the plentiful space-based solar power \cite{oltean_why_2024,wired2025-space-datacenters,arcas_towards_2025,wsj2025-space-datacenters}. Such an approach avoids the transmission losses between LEO and ground stations, but requires launching and constructing large-scale datacenters in orbit. 
The technical details have only been elaborated in two proposals: Starcloud \cite{oltean_why_2024} describes the on-orbit assembly of a monolithic structure upon which solar panels and compute nodes are mounted, while Google's Suncatcher proposes the creation of a distributed LEO datacenter, consisting of a cluster of many formation-flying satellites interconnected through inter-satellite links (ISLs) \cite{arcas_towards_2025}. We illustrate such a distributed datacenter cluster in Figure \ref{fig:intro_diagram}.

There are many tradeoffs to consider between monolithic and distributed space-based datacenter architectures. Some drawbacks of monolithic designs include the need for in-space assembly, difficulties associated with maneuvering and correctly orienting large space structures, and large thermal gradients across large monolithic spacecraft. On the other hand, distributed designs avoid many of the challenges incurred by large orbiting structures, but require the stable formation-flying of dense satellite clusters and establishment of reliable, high-data-rate inter-satellite links \cite{arcas_towards_2025}. Nonetheless, distributed LEO datacenters will benefit from economies of scale and technological advances that have enabled the recent proliferation of satellite constellations both in LEO and beyond Earth orbit \cite{michel_first_2022, penot_orbital_2026}.  

\begin{figure}
    \centering
    \includegraphics[width=0.42\linewidth]{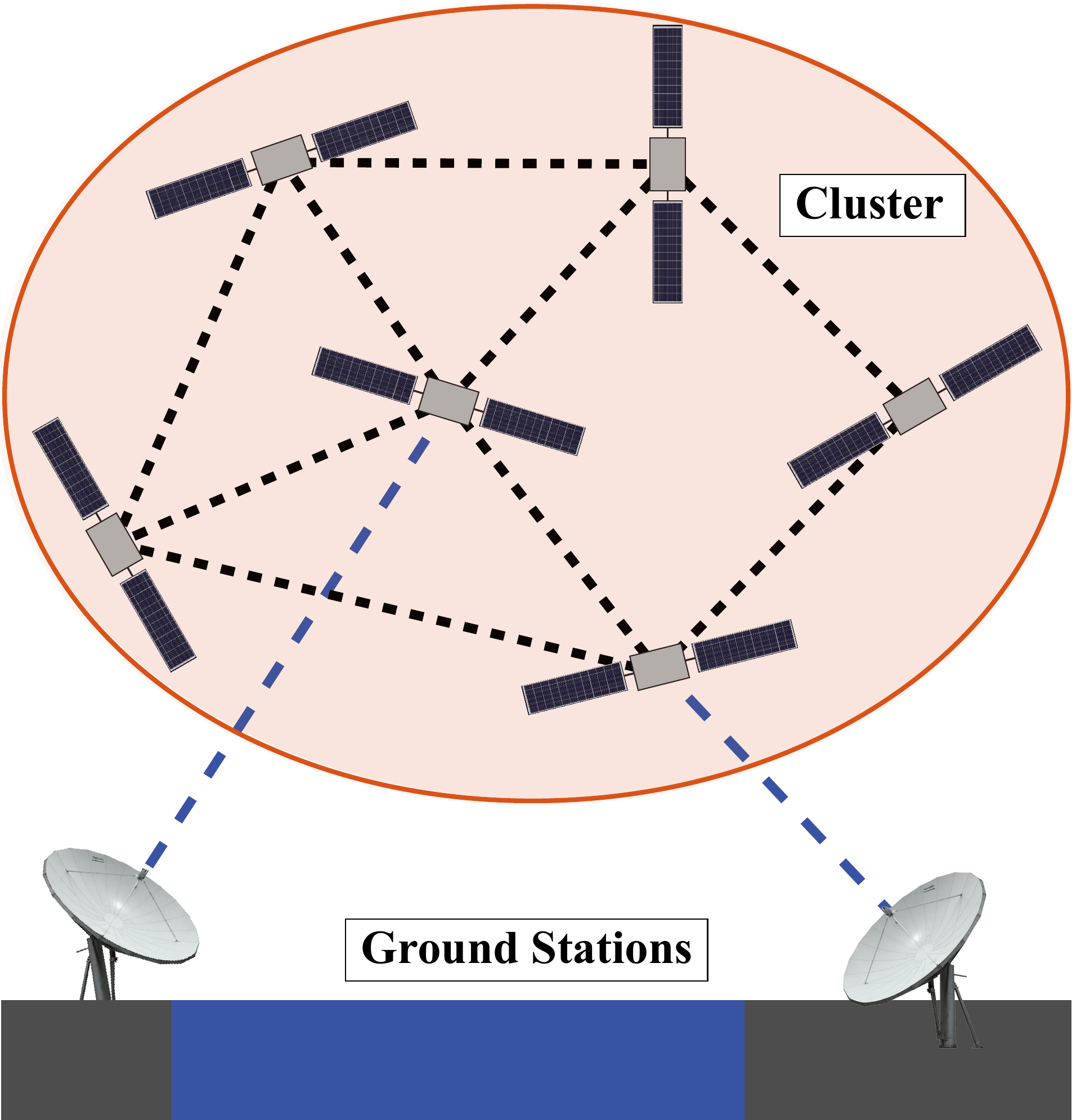}
    \caption{Diagram representing a distributed low Earth orbit (LEO) datacenter composed of multiple formation-flying satellites connected via inter-satellite links (ISLs). The components depicted in the figure are not to scale.}
    \label{fig:intro_diagram}
\end{figure}

In this work, we focus on the orbital design of the dense satellite clusters required in a distributed orbital datacenter. In designing such clusters, we conform to several safety and operational constraints, including enforcing minimum and maximum inter-satellite spacings, the availability of unobstructed ISLs, and continuous solar exposure for all satellite solar panels. Under these constraints, we develop and analyze two distinct cluster orbital designs: a planar design and a 3D design. We show that the former is the planar architecture that maximizes the number of cluster satellites $N_\text{sats}$ for given minimum and maximum inter-satellite spacings, yielding four times as many cluster satellites as the Suncatcher design, under the same constraints. The proposed 3D design allows $N_\text{sats}$ to scale proportionally to the cube of the cluster radius ($R_{\max}$), outperforming alternatives that scale proportionally to $R_{\max}^2$. Finally, we formulate and solve an integer optimization problem that maps a Clos VL2-like datacenter switching network onto a physical satellite cluster, showing that for both proposed cluster designs it is possible to replicate a high bisection-bandwidth, terrestrial datacenter-like network within the satellite cluster.

\section{Problem Statement}

In this work, we define and enforce several operational requirements relevant to the orbital design of LEO datacenters. These are primarily the same constraints as those defined in the Google Suncatcher white paper \cite{arcas_towards_2025}. These constraints apply to all designed clusters over their satellites' entire orbits.

\begin{enumerate}
    \item Collision avoidance: All cluster satellites must remain a minimum inter-satellite distance $R_{\min}$ away from other cluster satellites.
    \item Maximal inter-satellite distance: All cluster satellites must remain within a sphere of radius $R_{\max}$, guaranteeing a maximum inter-satellite distance of 2$R_{\max}$. 
    \item Maximizing solar exposure: All cluster satellites must have an unobstructed view of the Sun in order to keep their solar panels operating at 100\% capacity.
    \item Inter-satellite link stability: All cluster satellites must have a stable set of nearest neighbors with whom they can establish unobstructed inter-satellite links (ISLs). 
\end{enumerate}

\begin{figure} [h!]
    \centering
    \includegraphics[width=0.6\linewidth]{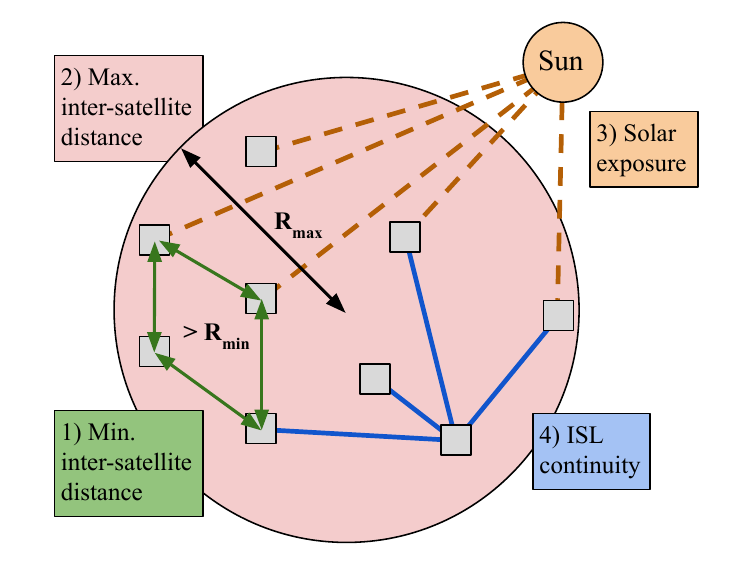}
    \caption{Diagram representing the LEO datacenter satellite cluster requirements. All four requirements apply to every cluster satellite.}
    \label{fig:cluster_reqs_diagram}
\end{figure}

The first two constraints, imposing an $R_{\min}$ and an $R_{\max}$, limit the number of satellites that can be packed into the cluster. Meanwhile, the solar exposure and ISL stability constraints limit the available cluster orbital geometries. 

In this work, we seek to \textbf{maximize the number of satellites} $N_\text{sats}$ that can fit into a cluster for a given pair of ($R_{\min}$,$R_{\max}$) parameters, all the while enforcing the solar exposure and ISL availability constraints. We verify conformity with the latter two constraints both by construction and numerical analysis, for a range of satellite obstruction radii $R_\text{sat}$. Although we derive all results as a function of ($R_{\min}$,$R_{\max}$), for ease of comparison with the Suncatcher satellite cluster design most plots will use the ($R_{\min}$,$R_{\max}$) values assumed in the Suncatcher white paper, $R_{\min} = $ 100~m and $R_{\max} =$ 1000~m.

\section{Orbital Design}

\subsection{Methodology}

In this work, we use a set of well-posed, mean relative orbital elements based on the relative orbital elements (ROEs)\cite{damico_autonomous_2010} as shown below:

\begin{equation}
\label{eq:damico_ROEs}
\delta \boldsymbol{\alpha}
=
\begin{pmatrix}
\delta a \\
\delta \lambda \\
\delta e_x \\
\delta e_y \\
\delta i_x \\
\delta i_y
\end{pmatrix}
=
\begin{pmatrix}
(a_d - a_c)/{a_c} \\
(M_d - M_c) + (\Omega_d - \Omega_c)\cos i_c \\
e_d \cos{\omega_d} - e_c \cos{\omega_c} \\
e_d \sin{\omega_d} - e_c \sin{\omega_c} \\
i_d - i_c \\
(\Omega_d - \Omega_c)\sin i_c
\end{pmatrix}
\end{equation}

where $\{a_d, e_d, i_d, \omega_d, \Omega_d, M_d \}$ and $\{a_c, e_c, i_c, \omega_c, \Omega_c, M_c \}$ are the Keplerian orbital elements of the deputy spacecraft and the chief spacecraft respectively, in the Earth-centered inertial (ECI) frame. 

LEO datacenters reside in sun-synchronous LEO; in this work, we use a cluster altitude $z =$ 650 km ($a_c=R_\text{E} + z = 7028$ km), an altitude at which a sun-synchronous orbit lies at an inclination $i_c =$ 98°. However, we design and analyze the satellite clusters in the cluster center's Hill frame, a rotating frame centered at the center of the cluster such that the -x vector always points to the center of the Earth, and the y unit vector is always aligned with the cluster's direction of orbital motion. We represent this frame for clarity in Figure \ref{fig:HillFrame_diagram}.

\begin{figure} [h!]
    \centering
    \includegraphics[width=0.5\linewidth]{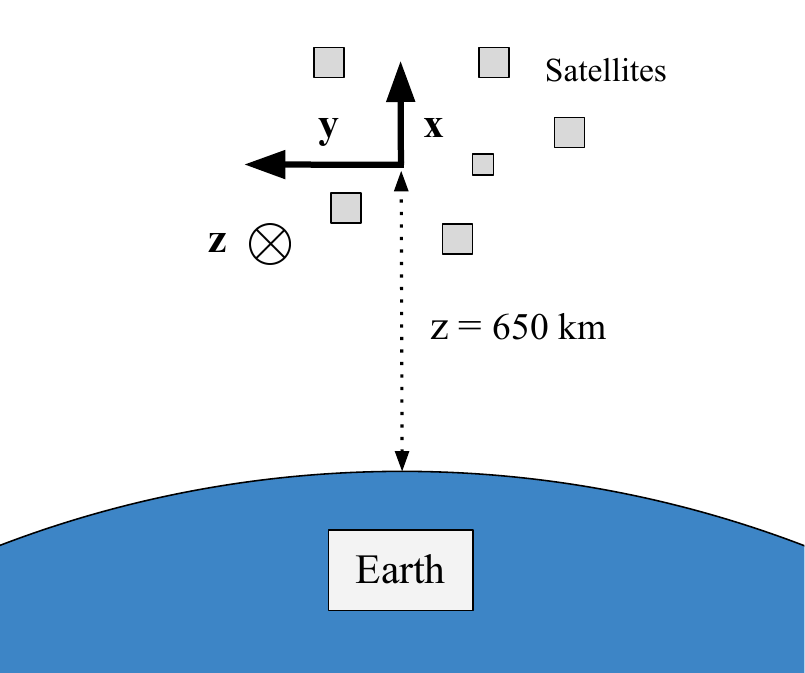}
    \caption{Diagram representing the Hill frame of the center of the cluster.}
    \label{fig:HillFrame_diagram}
\end{figure}

For convenience, we rotate the ECI frame such that an inclination of 0° corresponds to being in the plane of the satellite cluster's 98° sun-synchronous orbit. In our chosen frame, we therefore have $i_c=$ 0° and $e_c =$ 0, which causes $\omega_c$ and $\Omega_c$ to be undefined. Similarly, the ROE formulation in Equation (\ref{eq:damico_ROEs}) also runs into singularity issues when $i_d =$ 0° or $e_d =$ 0. In this work, we therefore use the following ROEs, modified such that they are non-singular for the types of satellite orbits used, and physically intuitive for designing dense satellite clusters.

\begin{equation}
\label{eq:our_ROEs}
\begin{pmatrix}
\delta a \\
\delta \lambda \\
\delta e_x \\
\delta e_y \\
\delta i_x \\
\delta i_y
\end{pmatrix}
=
\begin{pmatrix}
(a_d - a_c)/{a_c} \\
(M_d - M_c) + (\Omega_d - \Omega_c) + (\omega_d - \omega_c) \\
e_d \cos{\varpi_d} \\
e_d \sin{\varpi_d} \\
i_d \cos{\Omega_d} \\
i_d \sin{\Omega_d} 
\end{pmatrix}
\end{equation}

where $\varpi_d = \omega_d + \Omega_d$ is the longitude of perigee. 

Since all satellites in the LEO clusters in question must stay in close proximity over many orbits, they must have the same orbital period and therefore $a_d=a_c$. For all cluster satellites, we therefore have $\delta a= (a_d-a_c)/a_c = 0$ by construction.

While we use these modified ROEs to design and elicit some of the behaviors of the satellite clusters, we convert the ROEs to Keplerian orbital elements to model the evolution of the cluster structure over the course of its orbit around the Earth. The relationship between an elliptical orbit's mean anomaly $M$ and its true anomaly $\theta$ is given by the following transcendental equation, derived from Kepler's Equation:

\begin{equation}
    \label{eq:transcendentalEquation}
    M = \text{atan}2[\sqrt{1-e^2} \cdot \text{sin}(\theta), e + \text{cos}(\theta)] \ - \ \frac{e \cdot \sqrt{1-e^2} \cdot \text{sin}(\theta)}{1+ e \cdot\text{cos}(\theta)}
\end{equation}

To propagate the satellites' trajectories along their orbits, we propagate their mean anomaly $M$ linearly in time and solve Equation (\ref{eq:transcendentalEquation}) for the corresponding true anomaly $\theta$. At each timestep, we then convert from Keplerian to ECI frame Cartesian coordinates, and finally convert those into the satellites' Cartesian coordinates in the Hill frame portrayed in Figure \ref{fig:HillFrame_diagram} \cite{vallado_fundamentals_2013}.

\subsection{Suncatcher Orbital Design}

One of the major design constraints imposed upon the structure of satellite clusters for distributed LEO datacenters is that each satellite's solar panels be in sunlight throughout the cluster's entire orbit. In particular, this means that satellites should not shadow each other's solar panels. A natural way of meeting this solar exposure design constraint is to construct a satellite cluster wherein all satellites lie in a planar formation, as long as the Sun is always at some non-zero incident angle with respect to that plane. 

This is the approach taken in the Suncatcher orbital design, where the satellites are placed in elliptical mutually orbiting groups such that they lie in a planar rectangular grid \cite{arcas_towards_2025} ($\delta i_x = \delta i_y = i_d=$ 0°). However, this design is far from the optimal planar solution with regard to maximizing the number of satellites in the cluster for given values of $R_{\min}$ and $R_{\max}$. We identify the following contributing inefficiencies:

\begin{enumerate}
    \item Small planar mutually orbiting groups of argument of periapsis and mean anomaly produce satellite relative orbits of eccentricity $\sqrt{3}/2$ in the cluster center's Hill frame \cite{penot_heliocentric_2025}. This eccentricity limits the usable area of the $R_{\max}$-radius disk in which one can deploy satellites to that of the inscribed $\sqrt{3}/2$-eccentricity ellipse, as we plot in Figure \ref{fig:Suncatcher_orbital_design_plot}. 
    \item Since all satellites follow an elliptical relative orbit of eccentricity $\sqrt{3}/2$, their trajectories have a ratio of semi-major axis to semi-minor axis $a/b=2$. For a planar grid of satellites to respect a minimum inter-satellite spacing $R_{\min}$, when they cross the y-axis the satellites must be at an inter-satellite distance $\delta y = 2R_{\min} = 2 \delta x$, limiting the packing density of the planar formation.
    \item The Suncatcher cluster places its satellites in a rectangular lattice, while a hexagonal lattice is a more efficient packing of satellites onto a given plane. 
\end{enumerate}

\begin{figure} [h!]
    \centering
    \includegraphics[width=0.55\linewidth]{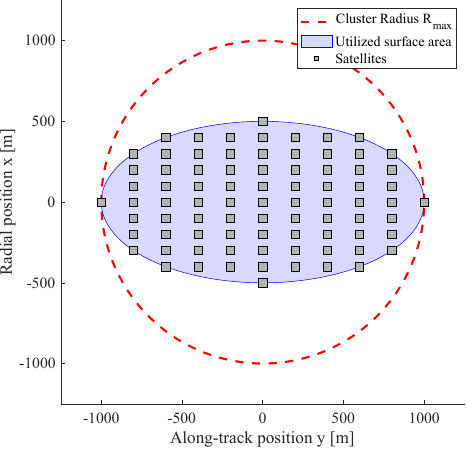}
    \caption{The Suncatcher orbital design fits $N_\text{sats} = $ 81 satellites into a cluster of $R_{\min}=$ 100~m and $R_{\max}=$ 1000~m \cite{arcas_towards_2025}.}
    \label{fig:Suncatcher_orbital_design_plot}
\end{figure}

We seek to remedy the above inefficiencies and design a $N_\text{sats}$-optimal planar cluster architecture as a function of ($R_{\min}$,$R_{\max}$). 

\subsection{Optimal Planar Solution}

We can obtain satellite relative orbits that are inclined in the Hill frame by giving the satellites non-zero $\delta i_x$ and $\delta i_y$ values. In particular, as long as all cluster satellites have the same ratio of $e_d/i_d$ and the same $\Omega_d$, then all satellites lie in the same plane, at some non-zero inclination $i_{\text{local}}$ with respect to the x-y plane of the Hill frame. Furthermore, if all satellite orbits are such that $\delta \lambda = 0$, this plane passes through the center of the cluster.

Varying $\Omega_d$ and $\varpi_d$ changes the orientation of the inclined plane. Let's define $\phi = \varpi_d + \Omega_d = \arctan(\delta e_y/ \delta e_x) + \arctan(\delta i_y / \delta i_x)$, the angle between the positive x-axis and the position of the largest positive displacement $z$ from the x-y plane. $\phi = k \pi, \forall k \in \mathbb{Z}$ yields planes that are inclined in the radial/cross-track direction, while $\phi = (k+\frac{1}{2}) \pi, \forall k \in \mathbb{Z}$ yields planes inclined in the along-track (y) direction. Other values of $\phi$ yield intermediate plane orientations. We plot the planes corresponding to some example $\phi$ values in Figure \ref{fig:inclined_planes_rotated}. 

\begin{figure} [h!]
    \centering
    \includegraphics[width=0.65\linewidth]{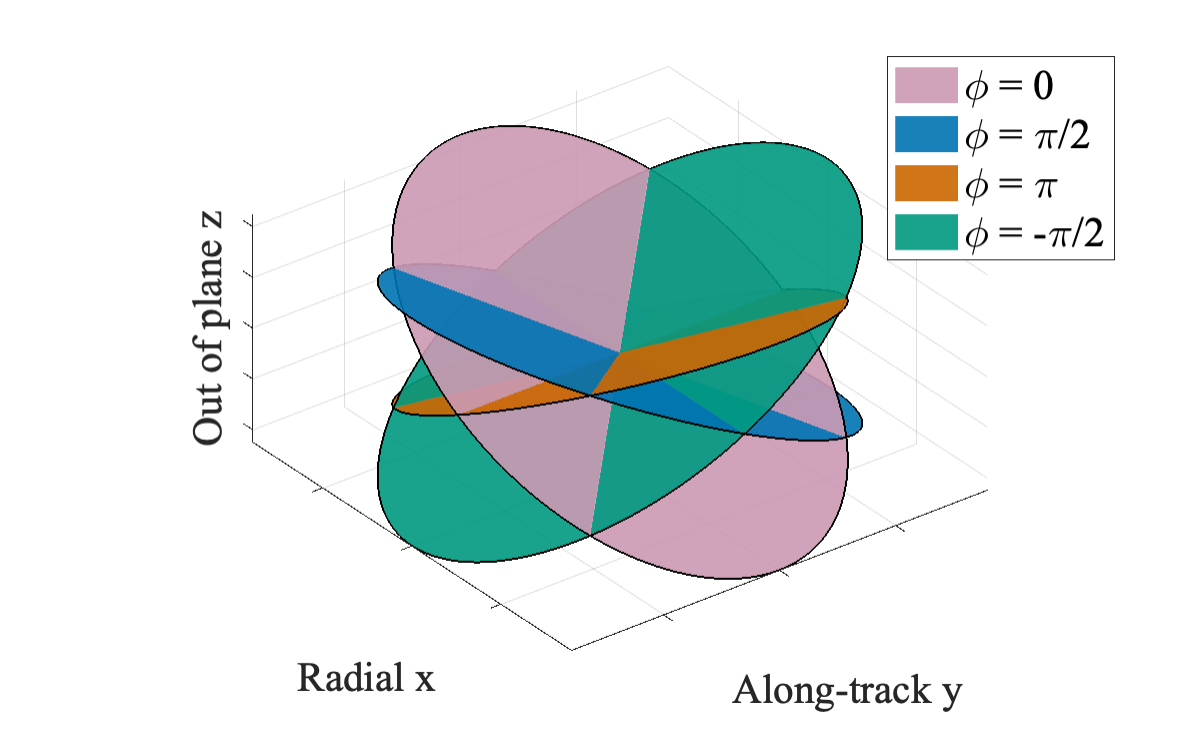}
    \caption{Plane orientation corresponding to several values of $\phi = \varpi_d + \Omega_d$.}
    \label{fig:inclined_planes_rotated}
\end{figure}

When the cluster plane is inclined in the radial direction, the minor axis of the satellites' elliptical relative orbits around the center of the cluster is extended, changing the aspect ratio and therefore the eccentricity of the satellite trajectories. If we set $i_d = \sqrt{3} \cdot e_d$, we obtain a local plane inclination $i_\text{local} = \pi /3$ radians and a circular trajectory for all satellites in the plane. This type of circular relative orbit is notably leveraged in a heliocentric setting by the proposed Laser Interferometer Space Antenna (LISA) mission's orbital design to place satellites in a circular trajectory around the center of the formation \cite{martens_trajectory_2021}.

Placing the satellites into circular trajectories around the center of the cluster has three main purposes:

\begin{enumerate}
    \item The planar satellite cluster can occupy the entire surface of the $R_{\max}$-radius disk.
    \item All trajectories are circular, so inter-satellite distances are fixed throughout the cluster's orbit. It is thus possible to pack satellites into an $R_{\min}$-spaced hexagonal lattice without violating the minimum inter-satellite spacing.
    \item The cluster rotates rigidly around the cluster's center, preserving the relative position of satellites and therefore guaranteeing unobstructed ISLs by construction.
\end{enumerate}

We plot the initial satellite positions of this planar cluster design in Figure \ref{fig:planarSol_3Dplot} for $R_{\min}=$ 100~m, $R_{\max}=$ 1000~m. We are able to place $N_\text{sats} =$ 367 satellites into the cluster with this optimal planar solution, more than four times as many as the original Suncatcher planar cluster represented in Figure \ref{fig:Suncatcher_orbital_design_plot}.

\begin{figure} [h!]
    \centering
    \includegraphics[width=0.7\linewidth]{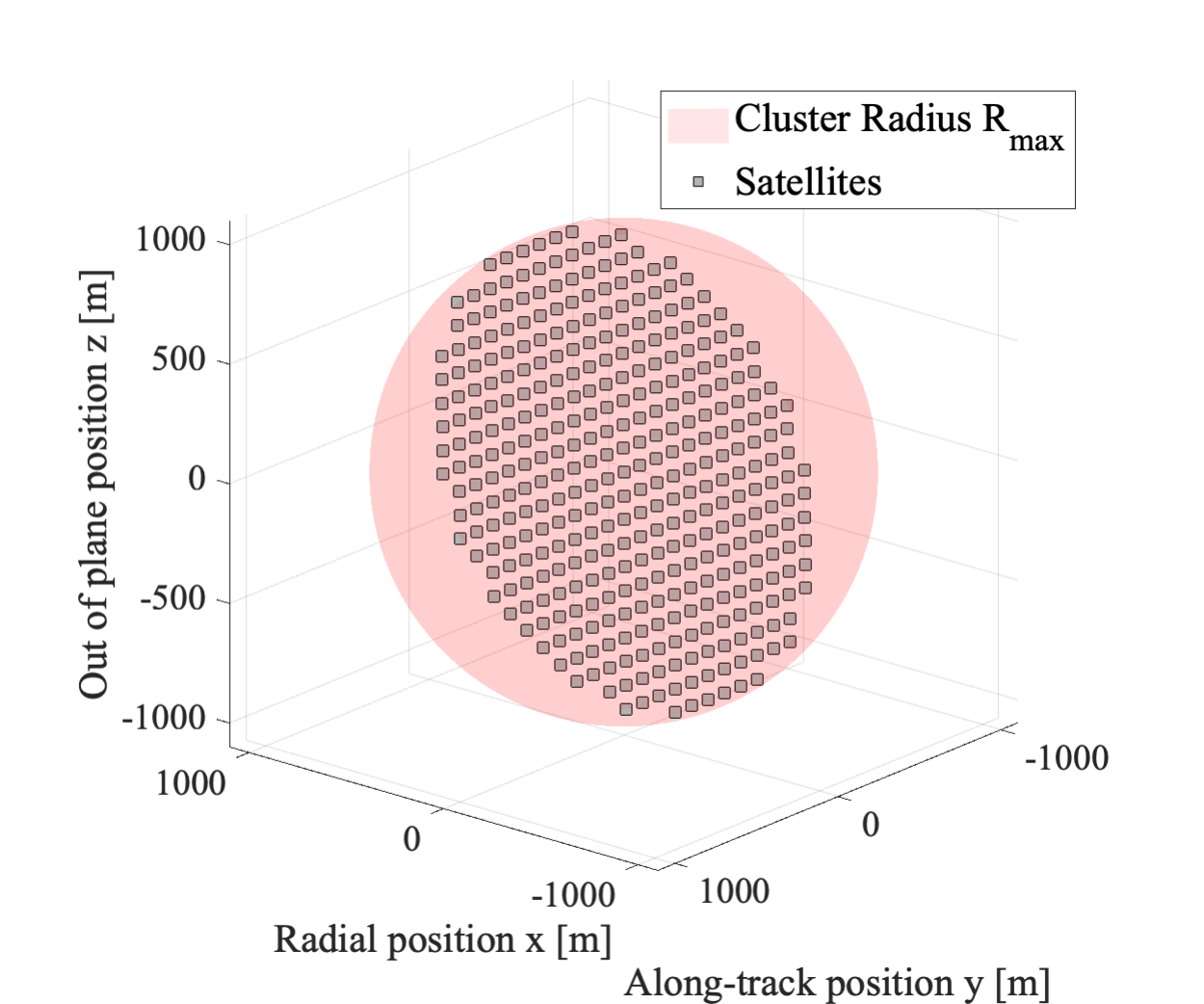}
    \caption{Optimal planar cluster design for $R_{\min}=$ 100~m, $R_{\max}=$ 1000~m, overlaid with a sphere of radius $R_{\max}$. With this cluster configuration, we are able to place $N_\text{sats} =$ 367 satellites.}
    \label{fig:planarSol_3Dplot}
\end{figure}

For given ($R_{\min}$, $R_{\max}$) values, this design approach yields the maximum possible $N_\text{sats}$ of any planar cluster. Indeed, it occupies a full cross-section of the $R_{\max}$ sphere, and tiles that surface area with a hexagonal lattice of inter-satellite spacing $R_{\min}$. Hexagonal tilings optimally pack a 2D plane subject to a constant inter-node minimum spacing \cite{grunbaum_tilings_1987}. For our circular planar cluster, the exact optimal tiling may or may not include a satellite at the center of the cluster, depending on the value of $R_{\max}/R_{\min}$. These planar clusters also comply with the solar exposure and stable ISL requirements by construction, as long as satellite dimensions aren't comparable to the inter-satellite spacing $R_{\min}$.

\subsection{3D Solution}

The cluster orbital design may be the optimal planar solution, but it doesn't utilize any of the volume of the limiting $R_{\max}$ sphere outside of the cluster plane. In this section, we develop a cluster orbital design approach that is no longer constrained to a single plane. Unlike our proposed planar design, not all possible 3D clusters comply with the solar exposure constraint by construction – we will instead verify compliance through numerical simulation in later sections.

Our approach to a dense 3D cluster is to stack inclined grids of satellites in the along-track (y) direction. This requires each plane within the cluster to be inclined in the along-track (y) direction, in other words $\phi = (k + \frac{1}{2})\cdot \pi, k \in \mathbb{Z}$. 

In such a cluster, each inclined plane has the same local inclination $i_\text{local}$, related to the magnitudes of each satellite's relative inclination and eccentricity vectors by the equation below:

\begin{equation}
    \label{eq:ilocal}
    i_\text{local} = \arctan{(\frac{2 \cdot i_\text{d}}{e_\text{d}})}
\end{equation}

All $R_{\min}$ constraints are preserved by construction within each inclined plane. To enforce a spacing of $R_{\min}$ between neighboring planes, we stagger the planes in their $\delta \lambda$ by an interval of $\Delta (\delta \lambda) = R_{\min}/(a_c \cdot \min{(\cos{(i_\text{local})},\sin{(i_\text{local})})})$. 

To construct the final formation, we therefore populate the cluster with $N_\text{planes} =  \lfloor R_{\max} / (a_c \cdot \Delta(\delta \lambda) )\rfloor + 1$ planes. We then propagate all satellites' trajectories over the course of an entire orbit of the cluster, pruning any satellites that stray out of the $R_{\max}$-radius outer sphere at any point in their trajectory. We plot radial, along-track, and out-of-plane views of an example 3D cluster for $R_{\min}=$ 100~m, $R_{\max}=$ 1000~m, and $i_\text{local} = $ 39°, in Figure \ref{fig:3Dsol_plots}.

For given ($R_{\min}$,$R_{\max}$) values, we optimize over $i_\text{local}$ in order to maximize $N_\text{sats}$. We plot $N_\text{sats}$ as a function of $i_\text{local}$ for example $R_{\min}$ and $R_{\max}$ values in Figure \ref{fig:NsatsVSilocal}. The maximal $N_\text{sats}$ value is typically attained for a range of $i_\text{local}$ values (41.2° to 43.8° in the example plotted), preserving a degree of freedom in the cluster design. We can therefore adjust $i_\text{local}$ to ensure compliance with the solar exposure design requirement, as we will discuss in a later section.

\begin{figure*}[h!]
    \centering
    \begin{subfigure}{0.32\textwidth}
        \centering
        \includegraphics[width=\linewidth]{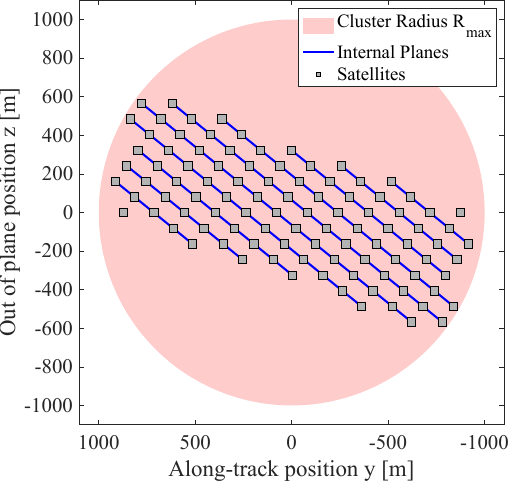}
        \caption{Radial view.}
        \label{fig:3Dsol_radialview}
    \end{subfigure}
    \hfill
    \begin{subfigure}{0.32\textwidth}
        \centering
        \includegraphics[width=\linewidth]{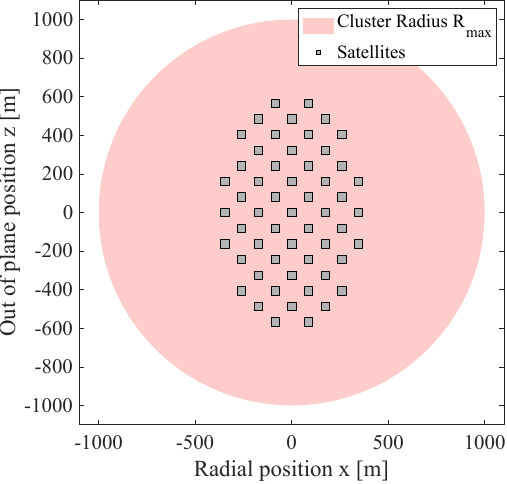}
        \caption{Along-track view.}
        \label{fig:3Dsol_alongtrackview}
    \end{subfigure}
    \hfill
    \begin{subfigure}{0.32\textwidth}
        \centering
        \includegraphics[width=\linewidth]{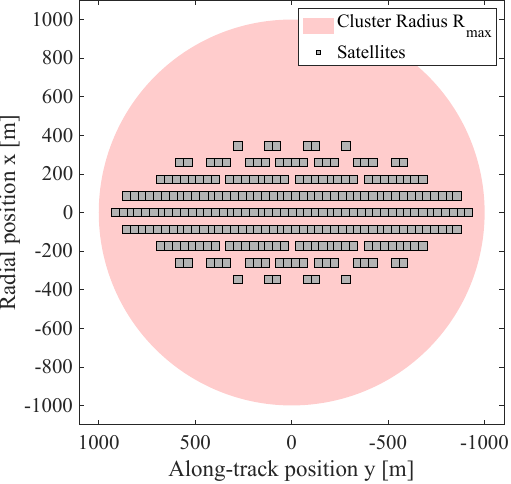}
        \caption{Out-of-plane view.}
        \label{fig:3Dsol_outofplaneview}
    \end{subfigure}
    \caption{3D cluster initial layout for $R_{\min}=$ 100~m, $R_{\max}=$ 1000~m, and $i_\text{local} = $ 39°. The radial view includes a visualization of the inclined planes of different $\delta\lambda$ that make up the 3D cluster. For these cluster parameters, $N_\text{sats} = $ 264 satellites, underperforming the optimal planar orbital design plotted in Figure \ref{fig:planarSol_3Dplot}.}
    \label{fig:3Dsol_plots}
\end{figure*}

\begin{figure} [h!]
    \centering
    \includegraphics[width=0.55\linewidth]{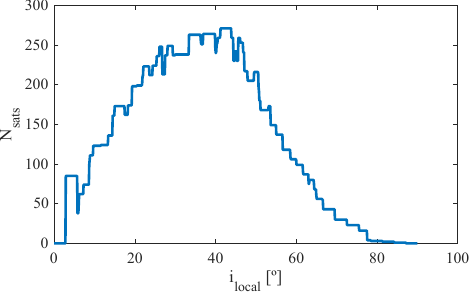}
    \caption{$N_\text{sats}$ vs. $i_\text{local}$ for $R_{\min} = $ 100~m and $R_{\max} = $ 1000~m.}
    \label{fig:NsatsVSilocal}
\end{figure}

\subsection[Nsats Scaling]{$N_\mathrm{sats}$ Scaling}

For each cluster orbital design, we compute the number of satellites $N_\mathrm{sats}$ deployable as a function of the ratio of $R_{\max}$ to $R_{\min}$, and plot the results in Figure \ref{fig:Nsats_scaling}. As expected, our proposed optimal planar solution outperforms the Suncatcher design for all $R_{\max}/R_{\min}$ ratios considered. Meanwhile, our proposed 3D design underperforms the planar design for small $R_{\max}/R_{\min}$, including at the original Suncatcher parameters ($R_{\min} = $ 100~m and $R_{\max} = $ 1000~m). The 3D design, however, scales better than the planar solutions, yielding larger $N_\text{sats}$ than the planar designs for $R_{\max}/R_{\min} \geq $ 13.5. 

\begin{figure} [h!]
    \centering
    \includegraphics[width=0.7\linewidth]{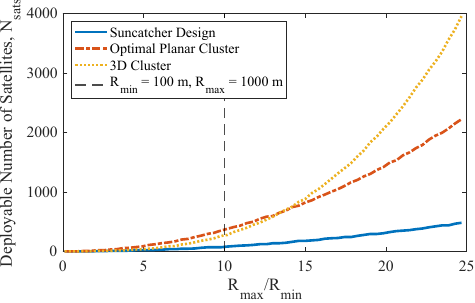}
    \caption{Number of cluster satellites $N_\text{sats}$ as a function of the ratio of $R_{\max}$ to $R_{\min}$, for the Suncatcher cluster design, the optimal planar cluster, and our proposed 3D cluster.}
    \label{fig:Nsats_scaling}
\end{figure}

We perform a power fit of the data plotted in Figure \ref{fig:Nsats_scaling} and report the coefficients and RMSE of each fit in Table \ref{tab:Nsats_scaling}. For a fixed inter-satellite spacing $R_{\min}$, both planar solutions scale approximately with $R_{\max}^2$, which is to be expected for clusters confined to a single plane. By contrast, our proposed 3D cluster design scales approximately with $R_{\max}^3$, outperforming both planar architectures for large datacenter satellite clusters.

\renewcommand{\arraystretch}{1.2}
\begin{table} [h!]
    \centering
    \begin{tabular}{|c|c|c|c|}
    \hline 
         Cluster Orbital Design & $a$ & $b$ & RMSE [sats] \\
         \hline
        Suncatcher Design & 0.80 & 1.996 & 4.65 \\
        Optimal Planar Cluster & 3.63 & 2.00 & 8.99 \\
        3D Cluster & 0.27 & 2.99 & 16.40 \\
        \hline
    \end{tabular}
    
    \vspace{2mm}
    \caption{Coefficients and RMSE of applying a power fit of the form $N_\text{sats} = a \cdot (R_{\max}/R_{\min})^b$ to the data plotted in Figure \ref{fig:Nsats_scaling}.}
    \label{tab:Nsats_scaling}
\end{table}

\section{Solar Exposure Analysis}

By construction, both the Suncatcher and our optimal planar orbital designs fulfill the solar exposure design requirement as we have formulated it: with no consideration of the actual dimensions of the cluster satellites and their solar panels. Even the proposed 3D cluster fulfills the solar exposure design requirement, if we model the satellites as points in space, as is evident from the out-of-plane view of the 3D cluster plotted in Figure \ref{fig:3Dsol_outofplaneview}. 

In this section, we conduct a more thorough analysis of the solar exposure of satellites in the different LEO datacenter clusters under consideration, by assuming that the satellites and their panels have a circular cross-section of non-zero radius $R_\text{sat}$. For all considered $R_\text{sat}$ values, we then numerically evaluate the average solar power output of each satellite as a fraction of its panels' total capacity.

\subsection{Sun Vector Model}

At our assumed orbital altitude $z = $ 650~km, the cluster itself must have an inclination of approximately $i_c =$ 98°. In the cluster Hill frame, this corresponds to the solar vector rotating 8° off of the z-axis (the out-of-plane direction) with a period equal to the cluster's orbital period $T_\text{cluster} = 2 \pi \cdot \sqrt{a_c^3/\mu_\text{E}}$, where $\mu_\text{E}$ is the Earth's gravitational parameter. 

In our solar exposure numerical analysis, we therefore model the sun vector at time $t$, $\vec{d}_\text{solar}(t)$, in the cluster Hill frame according to the equation below:

\begin{equation}
    \label{eq:sunvector_overtime}
    \vec{d}_\text{solar}(t) =  \begin{pmatrix}
        \cos(2 \pi \cdot t /T_\text{cluster}) \\
        \sin(2 \pi \cdot t /T_\text{cluster}) \\
        |\tan(i_c)|
    \end{pmatrix}
\end{equation}

\subsection{Solar Exposure}

In this analysis, we consider that each cluster satellite both obstructs and receives solar flux as if it were a disk of solar panels oriented towards the sun vector $\vec{d}_\text{solar}$ at all times. Deployed LEO datacenter satellites would likely have a different solar panel shape, but our circular approximation provides a useful estimate of the worst-case inter-satellite shadowing that could occur assuming all satellite hardware fits inside a sphere of radius $R_\text{sat}$ around the center of the satellite.

\subsection{Results}

\begin{figure} [h!]
    \centering
    \includegraphics[width=0.65\linewidth]{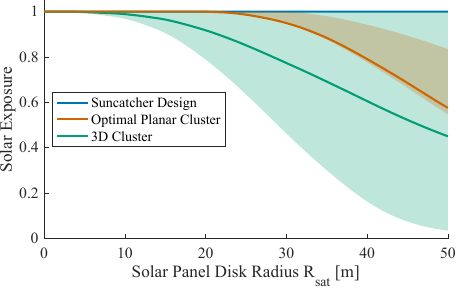}
    \caption{Solar exposure of cluster satellites as a function of $R_\text{sat}$ for $R_{\min}=$ 100~m and $R_{\max}=$ 1000~m for the three cluster designs under analysis. The 3D cluster was evaluated with $i_\text{local} = $ 43.8°. Solid lines plot the average exposure across the cluster, while the shaded areas represent the range of solar exposures across the satellites in the cluster.}
    \label{fig:solarExposure_vs_Rsat}
\end{figure}

We conduct the solar exposure analysis for all three cluster orbital designs for $R_{\min}=$ 100~m, $R_{\max}=$ 1000~m, and a range of $R_\text{sat}$ values. We plot the resulting average solar exposure fraction in Figure \ref{fig:solarExposure_vs_Rsat}. The Suncatcher orbit design maintains full solar exposure for all its satellites for $R_\text{sat} < $ 50~m $= R_{\min}/2$. Meanwhile, for $R_\text{sat} \geq $ 19~m, our proposed planar cluster design also results in some inter-satellite solar occlusion. 

\begin{figure} [h!]
    \centering
    \includegraphics[width=0.65\linewidth]{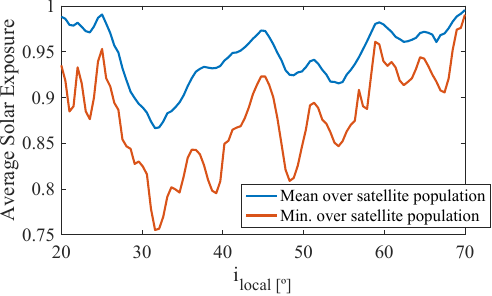}
    \caption{Solar exposure fraction averaged across the 3D cluster's orbit, as a function of $i_\text{local}$, for the proposed 3D architecture with $R_{\min}=$ 100~m, $R_{\max}=$ 1000~m, and $R_\text{sat} = $ 15~m.}
    \label{fig:solarExposure_vs_ilocal}
\end{figure}

For the 3D cluster with $R_{\min}=$ 100~m and $R_{\max}=$ 1000~m, we found that solar occlusion between satellites starts to occur for $R_\text{sat} \geq $ 3~m for all practical $i_\text{local}$ values. To select an $i_\text{local}$ value for the cluster, we evaluate the average and worst time-averaged solar exposure as a function of $i_\text{local}$, for $R_\text{sat} = $ 15~m (corresponding to half of the wingspan of a Starlink V2-Mini satellite). We plot those results in Figure \ref{fig:solarExposure_vs_ilocal}, where we observe that both the average and worst-satellite solar exposure increase monotonically with $i_\text{local}$ in our $N_\text{sats}$-optimal range of $i_\text{local} \in $ (41.2°, 43.8°). We therefore select $i_\text{local} = $ 43.8° as the plane inclination of the proposed 3D cluster to further analyze. As plotted in Figure \ref{fig:solarExposure_vs_Rsat}, the range of solar exposures is much larger for the 3D cluster than for our proposed planar cluster: at $R_\text{sat} = $ ~50~m, some 3D cluster satellites still have a solar exposure near 100\%, while others are almost permanently shadowed.

\section{Cluster ISL Network Analysis}

In order to function like a terrestrial datacenter, on-orbit datacenters must be densely interconnected through ISLs. A naive approach would be to connect all satellites in a regular repeating mesh: in the case of our proposed planar cluster, a hexagonal mesh, and in the case of our proposed 3D cluster, a 3D lattice where each node connects to its 8 nearest neighbors. For a range of $R_{\max}$ (and resulting $N_\text{sats}$) values, we perform a spectral analysis of the cluster network graph and report the results in Table \ref{tab:mesh_scaling}.

\renewcommand{\arraystretch}{1.6}
\begin{table} [h!]
    \centering
    \begin{tabular}{|c|c|c|}
    \hline
    Parameter Name & Planar Cluster Scaling & 3D Cluster Scaling \\
    \hline
    Diameter & $\propto (N_\text{sats})^{\frac{1}{2}}$  & $\propto (N_\text{sats})^{\frac{1}{3}}$  \\
    \hline
    Mean Path Length & $\propto (N_\text{sats})^{\frac{1}{2}}$ & $\propto (N_\text{sats})^{\frac{1}{3}}$ \\
    \hline
    Bisection Bandwidth & $\propto (N_\text{sats})^{\frac{1}{2}}$ & $\propto (N_\text{sats})^{\frac{2}{3}}$ \\
    \hline
    Fiedler value, $\lambda_2$ & $\propto (N_\text{sats})^{-1}$ & $\propto (N_\text{sats})^{-\frac{2}{3}}$ \\
    \hline
    \end{tabular}
    \caption{Scaling of mesh internal cluster network for the proposed planar and 3D LEO datacenter cluster topologies.}
    \label{tab:mesh_scaling}
\end{table}

As expected, connecting cluster satellites in a simple mesh doesn't scale well with an increasing number of nodes $N_\text{sats}$. For clusters of at least a few dozen satellites, such network configurations would lead to excessive mean internode path lengths – furthermore, ISLs near the center of the formation would be congested due to their position in the cluster network graph, leading to further issues related to load balancing and congestion. Such a mesh network would thus not lend itself well to large-scale LEO datacenters.

\subsection{Clos Switching Network}

Another approach to designing the ISL network structure for a given LEO datacenter cluster is to replicate the structure of the switching fabric of terrestrial datacenters. 

Specifically, if we dedicate a subset of a cluster's satellites to being network switches (maintaining a large number of ISLs), while the rest of the satellites primarily carry compute hardware, we can recreate a VL2-like Clos switching network where compute satellites play the role of top-of-rack (ToR) switches, while the switching satellites play the role of datacenter network switches (both aggregation (AGG) and intermediate (INT) switches). 

First, we develop a method of designing a Clos network with $L$ layers of nodes, and an even number of ports $k$ at each non-ToR switch. $L = $ 1 corresponds to a small network of at most $k+1$ ToR nodes, where all nodes are interconnected. $L = $ 2 corresponds to a network of at most $k$ ToR nodes, each connected to every one of the $k/2$ INT nodes.

We base the Clos topology for $L =$ 3 on that defined in the VL2 paper \cite{greenberg_vl2_2009}. Each ToR switch connects to two AGG switches, while each AGG switch connects to every INT switch. We plot an example of such a Clos network in Figure \ref{fig:Clos_Lis3}. For $L \geq $ 4, we add an additional layer of aggregation nodes compared to the network structure at $L-1$. We present the resulting formulas for the total number of nodes and the number of ToR nodes supported at different numbers of layers $L$ in Table \ref{tab:ClosNetwork_nodeNs}. 

\renewcommand{\arraystretch}{1.5}
\begin{table}
    \centering
    \begin{tabular}{|c|c|c|}
    \hline
    $L$       & Max. Number of Nodes & Max. Number of ToR Nodes \\
    \hline
    1         & $k+1$ & $k+1$ \\
    \hline
    2         & $3k/2$  & $k$ \\
    \hline
    $\geq$ 3  & $(k/2)^{L-1} + (2L-3) (k/2)^{L-2}$ & $(k/2)^{L-1}$ \\
    \hline
    \end{tabular}
    \caption{Maximum total number of nodes, and maximum number of ToR nodes, supported for an $L$-layer, $k$-port Clos datacenter switching network.}
    \label{tab:ClosNetwork_nodeNs}
\end{table}

We plot example Clos networks for $L = $ 2, 3, and 4 in Figure \ref{fig:ClosNetwork_examples}. 

\begin{figure*}[h!]
    \centering
    \begin{subfigure}{0.32\textwidth}
        \centering
        \includegraphics[width=\linewidth]{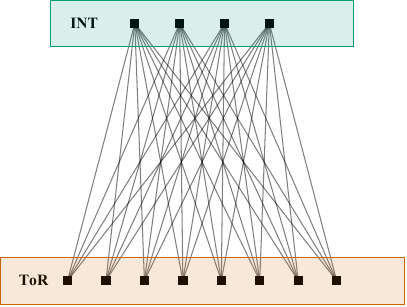}
        \caption{$L = $ 2}
        \label{fig:Clos_Lis2}
    \end{subfigure}
    \hfill
    \begin{subfigure}{0.32\textwidth}
        \centering
        \includegraphics[width=\linewidth]{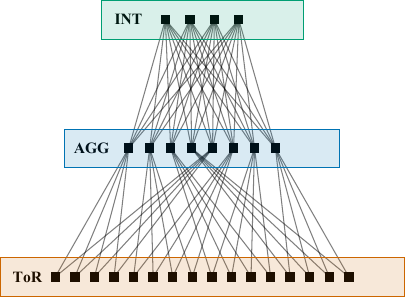}
        \caption{$L = $ 3}
        \label{fig:Clos_Lis3}
    \end{subfigure}
    \hfill
    \begin{subfigure}{0.32\textwidth}
        \centering
        \includegraphics[width=\linewidth]{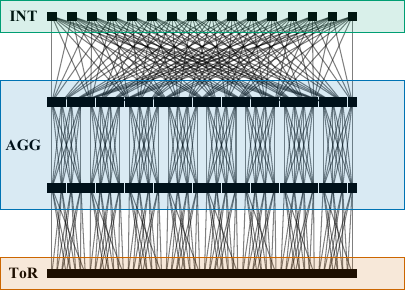}
        \caption{$L = $ 4}
        \label{fig:Clos_Lis4}
    \end{subfigure}
    \caption{Example Clos network topologies for $k = $ 8, for $L = $ 2, 3, and 4. We remove ToR and AGG nodes from these full topologies to adapt them to a particular cluster with $N_\text{sats}$ satellites.}
    \label{fig:ClosNetwork_examples}
\end{figure*}

For a cluster with $N_\text{sats}$ satellites, and an assumed maximum of $k$ ISLs per switching satellite, we use the formulas in Table \ref{tab:ClosNetwork_nodeNs} to find the required number of layers $L$. We then generate a Clos topology with $N_\text{sats}$ total nodes, pruning ToR and AGG nodes from the maximal L-layer network while maintaining full bisection bandwidth between all remaining ToR nodes.

\subsection{Integer Programming Formulation of Node Assignment}

We define the line of sight (LOS) matrix of a particular $N_\text{sats}$-satellite cluster as the $N_\text{sats}$-by-$N_\text{sats}$ binary matrix that indicates whether line of sight between any pair of cluster satellites is preserved throughout the entire cluster's orbit. For a given cluster, we precompute LOS numerically through a discrete simulation of the satellites' relative motion in the Hill frame, similar to our solar exposure analysis. At every timestep, we check whether each potential ISL passes within $R_\text{sat}$ from any uninvolved satellite in the cluster. $\text{LOS}(i,j) =$ 1 indicates that it is possible to maintain a continuous, unobstructed ISL between cluster satellites $i$ and $j$ over the entire course of the satellites' orbits.

To construct such a Clos network within a particular LEO datacenter cluster, we assign each physical satellite a corresponding virtual node in the generated Clos network graph, subject to the LOS constraints. We formulate this assignment problem as an integer program (IP), which we present in Equation (\ref{eq:Clos_IP_formulation}). The two equality constraints ensure that every physical satellite is assigned to exactly one virtual node, and each virtual node to exactly one physical satellite. The inequality constraints enforce the LOS constraints by forbidding pairs of node assignments that would require LOS-infeasible ISLs.

We define our decision variable $x_{ip}$ as follows:

\begin{equation}
    x_{ip} = 
      \begin{cases}
          1 \quad \text{if virtual node } i \text{ maps to satellite } p \\
          0 \quad \text{otherwise}
      \end{cases}
\end{equation}

We then obtain an integer optimization problem subject to the following constraints:

\begin{equation}
\label{eq:Clos_IP_formulation}
\begin{aligned}
    \text{find } \vec{x} \text{ s.t.} \\
    \sum_{i=1}^{N_\text{sats}} x_{ip} &= 1 \quad \forall \ p \in \{1,2,...,N_\text{sats}\} \\
    \sum_{p=1}^{N_\text{sats}} x_{ip} &= 1 \quad \forall \ i \in \{1,2,...,N_\text{sats}\} \\
    x_{ip}+x_{jq} &\leq 1
    \qquad
    \forall\, \{i,j\}\in E_\text{Clos},\;
    \forall\, p,q:\operatorname{LOS}(p,q)=0 \\
    \vec{x} &\in \{0,1\}^{N_\text{sats}^2}
\end{aligned}
\end{equation}

Without an objective function, this optimization problem is a feasibility check, returning the first feasible solution $\vec{x}$ encountered. We solve the integer optimization problem using Gurobi through its MATLAB API \cite{gurobi}.

We solve this optimization problem for both proposed clusters. We find that the assignment problem is often infeasible when $L \leq 2$, i.e. when $N_\text{sats} \leq 3k/2$, because the $L=$ 1 and $L =$ 2 Clos networks are very densely interconnected compared to the $L \geq $ 3 Clos networks. For $L \geq 3$, i.e. for $N_\text{sats} > 3k/2$, for both the optimal planar cluster and our proposed 3D cluster we found feasible solutions over the entire parameter ranges presented in Table \ref{tab:IP_parameter_ranges}.

\begin{table} [h!]
    \centering
    \begin{tabular}{|c|c|c|}
    \hline
    Parameter & Evaluation Range & Increment Size \\
    \hline
    $R_{\min}$ & 100~m & N/A \\
    \hline
    $R_{\max}$ & [500~m – 2000~m] & 100~m \\
    \hline
    k     & [4 – 12] & 2 \\
    \hline
    $R_\text{sat}$ & [0~m – 15~m] & 5~m \\
    \hline
    \end{tabular}
    \caption{Parameter ranges and intervals over which the integer program was solved for $L \geq 3$, for both proposed LEO datacenter cluster designs.}
    \label{tab:IP_parameter_ranges}
\end{table}

We plot example physical ISL solutions for the optimal planar cluster and the proposed 3D cluster in Figures \ref{fig:IP_planar_example_sol} and \ref{fig:IP_3D_example_sol}, respectively.

\begin{figure}
    \centering
    \includegraphics[width=0.7\linewidth]{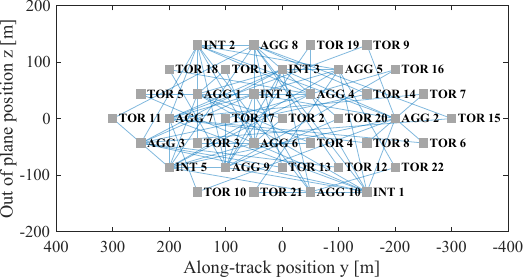}
    \caption{Physical ISL solution for the optimal planar cluster with $R_{\min} = $ 100~m, $R_{\max} = $ 300~m, $k = $ 10, and $R_\text{sat} = $ 15~m, viewed from the x-direction. The resulting cluster has $N_\text{sats} = $ 37 satellites and $L = $ 3 layers.}
    \label{fig:IP_planar_example_sol}
\end{figure}

\begin{figure}
    \centering
    \includegraphics[width=0.7\linewidth]{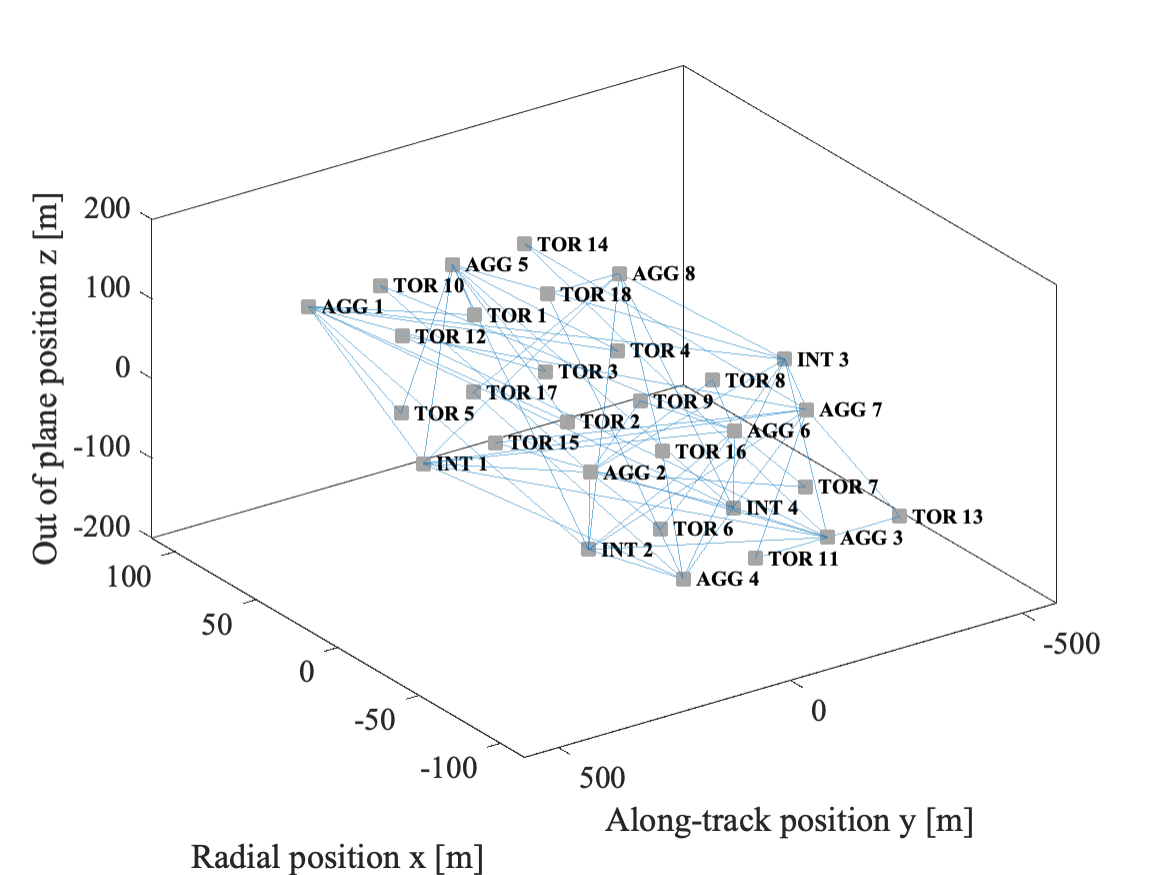}
    \caption{Physical ISL solution for the 3D cluster with $R_{\min} = $ 100~m, $R_{\max} = $ 500~m, $k = $ 10, and $R_\text{sat} = $ 15~m. The resulting cluster has $N_\text{sats} = $ 30 satellites and $L = $ 3 layers.}
    \label{fig:IP_3D_example_sol}
\end{figure}

\subsection{Clos Network Scaling}

One of the main drawbacks of constructing a VL2-like Clos network structure within a LEO datacenter satellite cluster is that such an approach requires some satellites to act as dedicated network switches. Due to their increased number of ISLs and therefore heavier communication hardware, it is reasonable to assume that said switching satellites would carry little or no compute hardware, reducing the total compute of the cluster compared to simpler inter-satellite network designs without dedicated switching nodes.

To mitigate this drawback, we identify the optimal number of Clos network layers $L$ and number of ISL links per switching satellite $k$ for a cluster with $N_\text{sats}$ satellites. Consider the fraction of cluster satellites $r(k,L)$ devoted to being compute satellites (ToR nodes) for a full Clos network with $L \geq 3$, as defined in Table \ref{tab:ClosNetwork_nodeNs}:

\begin{equation}
    \label{eq:rKL}
    r(k,L) = \frac{\text{max \# of ToR nodes}}{\text{max total \# of nodes}} = \frac{k}{k+4L-6}
\end{equation}

By inspection, $r(k,L)$ decreases with increasing $L$ but increases with increasing $k$. Let us suppose that there is some practical upper limit $k_{\max}$ on the number of simultaneous ISLs that a switching satellite can sustain (e.g. due to mass and power constraints). If $N_\text{sats} \leq k_{\max}+1$, then no switching nodes are required and $L=1$ is optimal. If $k_{\max} + 1 < N_\text{sats} \leq 3k_{\max}/2$, then $L=$ 2 is optimal. Otherwise, cluster designers should select the smallest number of layers $L_{\min}$ that satisfies the following capacity equation (adapted from Table \ref{tab:ClosNetwork_nodeNs}): 

\begin{equation}
    (k_{\max}/2)^{L_{\min}-1} + (2L_{\min}-3) (k_{\max}/2)^{L_{\min}-2} \geq N_\text{sats} 
\end{equation}

\section{Viability of LEO Datacenters}

It is important to acknowledge that significant questions remain unanswered regarding the practical viability of space-based datacenters. Challenges include the launch costs associated with satellite deployment, the impact of the radiation environment on datacenter hardware, the large data rates required between satellites, and the relative costs of constructing and operating space-based vs. terrestrial datacenters. While these questions were outside the scope of this paper, we have shown that it is possible to design dense satellite clusters that can meet the connectivity requirements that have been the basis of terrestrial datacenter architectures. 

\section{Conclusions}

Distributed space-based datacenters take advantage of the plentiful access to solar power in dawn-dusk sun-synchronous LEO, drawing significant attention and investment in academia and industry alike. However, existing literature is short on the orbital design of the dense satellite clusters needed to make such distributed LEO datacenters a reality.

In this work, we proposed two distinct LEO datacenter cluster orbital designs: a planar architecture, and a 3D architecture. We showed through construction and numerical analysis that both proposed architectures comply with relevant LEO datacenter design requirements, including guaranteeing a minimum inter-satellite spacing $R_{\min}$, a cluster radius $R_{\max}$, unobstructed solar power for all cluster satellites, and that each satellite have a stable set of nearest neighbors with which it can maintain ISLs. 

We demonstrated that our proposed planar orbital design is the planar cluster design that maximizes the number of deployable satellites $N_\text{sats}$ for any given $R_{\min}$ and $R_{\max}$ values, yielding $N_\text{sats}$ values four times those of our baseline, the Suncatcher orbital design. Furthermore, our proposed 3D design yields an $N_\text{sats}$ proportional to $(R_{\max}/R_{\min})^3$, scaling more rapidly than both planar orbital designs. Our $N_\text{sats}$-optimal planar cluster also rotates as if it were a rigid structure, greatly simplifying the design and establishment of permanent ISLs between satellites. We summarize the comparison of the different LEO datacenter orbital designs in Table \ref{tab:conclusion_summary}.

\begin{table} [h!]
    \centering
    \begin{tabular}{|c|c|c|c|}
    \hline
    Cluster Design & $N_\text{sats}$ Scaling & Full Solar Exposure & Rigid Rotation \\
    \hline
    Suncatcher & $\propto (R_{\max}/R_{\min})^2$ & $R_\text{sat} <$ 50~m & No \\
    \hline
    Optimal Planar & $\propto (R_{\max}/R_{\min})^2$ & $R_\text{sat} <$ 19~m & Yes \\
    \hline
    3D Cluster & $\propto (R_{\max}/R_{\min})^3$ & $R_\text{sat} <$ 3~m & No \\
    \hline
    \end{tabular}
    \caption{Comparison between the Suncatcher cluster design and the two cluster designs proposed in this paper.}
    \label{tab:conclusion_summary}
\end{table}

Finally, we formulated and solved an integer optimization problem to map VL2-like Clos network nodes to physical cluster satellites. We thus showed that it is possible to replicate a datacenter switching fabric within both proposed clusters for a wide range of $R_{\max}$ values and Clos network parameters, enabling the scaling of LEO datacenters to large numbers of satellites with similar load balancing and congestion avoidance properties as terrestrial datacenters. 

The proposed LEO datacenter orbital designs significantly improve upon the literature's baseline in terms of the number of satellites deployable into a given cluster radius. Furthermore, our Clos network to physical ISL mapping directly enables the reproduction of large-scale terrestrial datacenter switching topologies in orbit. Together, these results establish a foundation for the orbital and network co-design of large-scale space-based computing infrastructure.

\section{Acknowledgments}
The NASA University Leadership Initiative (grant \#80NSSC\allowbreak25M7102) and NASA grant \#80NSS{\allowbreak}C23M0220 provided funds to assist the authors with their research, but this article solely reflects the opinions and conclusions of its authors and not any NASA entity. J. Pénot was also supported in part by the National Science Foundation Graduate Research Fellowship under Grant No. 2024377105. Any opinions, findings, conclusions, or recommendations expressed in this material are those of the authors and do not necessarily reflect the views of the National Science Foundation.

Large language models were used solely for language editing and proofreading; all technical content, analyses, results, and conclusions are the authors’ own.

\bibliographystyle{AAS_publication}
\bibliography{references}

\end{document}